\documentclass[10pt]{article}
\usepackage{graphicx}
\usepackage{multirow}
\usepackage{subfigure}

\def\Title#1{\begin{center} {\Large #1 } \end{center}}
\def\Author#1{\begin{center}{ \sc #1} \end{center}}
\def\Address#1{\begin{center}{ \it #1} \end{center}}

\newcommand\pubblock{\rightline{\begin{tabular}{l} Proceedings of the Second Annual LHCP\\ \pubnumber\\
         \pubdate  \end{tabular}}}

\newenvironment{Abstract}{\begin{quotation} \begin{center} 
             \large ABSTRACT \end{center}\bigskip 
      \begin{center}\begin{large}}{\end{large}\end{center} \end{quotation}}

\newenvironment{Presented}{\begin{quotation} \begin{center} 
             PRESENTED AT\end{center}\bigskip 
      \begin{center}\begin{large}}{\end{large}\end{center} \end{quotation}}

\def\beq{\begin{equation}}
\def\eeq#1{\label{#1}\end{equation}}
\def\eeqn{\end{equation}}

\def\beqa{\begin{eqnarray}}
\def\eeqa#1{\label{#1}\end{eqnarray}}
\def\eeqan{\end{eqnarray}}

\let\bar=\overbar

\def\Dslash{\not{\hbox{\kern-4pt $D$}}}
\def\dslash{\not{\hbox{\kern-2pt $\del$}}}

\def\msb{{\bar{\ssstyle M \kern -1pt S}}}

\usepackage{color}

 \newcommand\pubnumber{ }

\newcommand\pubdate{\today}

\def\affiliation{
On behalf of the CMS Collaboration, \\
Lauritsen Laboratory of Physics \\
California Institute of Technology, Pasadena, CA 91125, U.S.A }

\begin{document}

% large size for the first page
\large
\begin{titlepage}
\pubblock

%% Change the title, name, abstract
%% Title 
\vfill
\Title{Search for natural Supersymmetry in events with 1 b-tagged jet
  using razor variables at $\sqrt{s}=8$ TeV}
\vfill

%  if you need to add the support use this, fill the \support definition above. 
%   \Author{ FIRSTNAME LASTNAME \support }
\Author{ Javier Duarte  }
\Address{\affiliation}
\vfill
\begin{Abstract}
We discuss a search for natural Supersymmetry in events with at least
one bottom-quark jet using razor variables in proton-proton
collisions at $\sqrt{s}=8$ TeV with the CMS detector. The event distribution in the
 plane defined by the razor variables $\mathrm{R^2}$ and
 $\mathrm{M_R}$ is studied, searching for a peaking signal on top of
 a smoothly falling standard model background. The data are consistent
 with the expected background, modeled by a template
 function. The 95\%~C.L. exclusion limit on the masses of the gluino and lightest supersymmetric
particle in a benchmark simplified model is presented. For a lightest
supersymmetric particle mass of $100$~GeV, the pair production of gluinos in
a multi-bottom final state is excluded for gluino masses up to $1375$~GeV.
\end{Abstract}
\vfill

% DO NOT CHANGE 
\begin{Presented}
The Second Annual Conference\\
 on Large Hadron Collider Physics \\
Columbia University, New York, U.S.A \\ 
June 2-7, 2014
\end{Presented}
\vfill
\end{titlepage}
\def\thefootnote{\fnsymbol{footnote}}
\setcounter{footnote}{0}
%

% normal size for the rest
\normalsize 

%% Your paper should be entered below. 

\section{Introduction}
R-parity conserving, weak-scale supersymmetry (SUSY) is a
well-motivated theory, which provides a suitable dark matter candidate
and predicts events at the LHC with jets and large missing transverse
momentum $\mathrm E_{\mathrm T}^{\mathrm{miss}}$. Natural SUSY models contain a light chargino
$\tilde\chi^{\pm}$ and a neutralino $\tilde\chi^0$ nearly degenerate in
mass, a light top or a bottom squark ($\tilde t$ or $\tilde b$), and potentially a
slightly heavier gluino $\tilde g$ in order to minimize the
fine-tuning associated with the observed value of the Higgs boson mass.

We discuss a search for squarks and gluinos in the
context of natural SUSY spectra, performed on events with two or more
jets, at least one of which is identified as originating from a bottom
quark~\cite{razor8TeV1,razor8TeV2}. The search is carried out on the data collected
by the Compact Muon Solenoid (CMS) Collaboration in proton-proton
collisions at $\sqrt{s}=8$ TeV in 2012, corresponding to an
integrated luminosity of 19.3 fb$^{-1}$. A complete description of the
CMS detector is given in~\cite{Chatrchyan:2008aa}. We utilize the razor
kinematic variables $\mathrm{R^2}$ and
$\mathrm{M_R}$~\cite{rogan,razor2010} to search for a broadly peaking
signal on the smoothly falling standard model (SM) background. The
analysis is performed in several disjoint datasets (referred to as
{\it boxes}), differing in the lepton and the jet multiplicity. In the
following, we discuss the results relating to the zero-lepton
boxes, which are analyzed in exclusive b-tagged jet
multiplicity bins, to maximize the sensitivity to direct and cascade
production of third generation squarks. This search extends a previous
analysis by CMS, performed with the same technique on the data
collected at a center-of-mass energy of
7~TeV~\cite{razorPRL,razorPRD}. 

\section{Razor variables and event selection}
In the canonical two-jet topology resulting from the production of two
squarks each decaying to a quark and the lightest SUSY particle
(LSP), the razor variables $\mathrm{M_R}$ and $\mathrm{R^2}$ are intended to
characterize the mass scale of the of SUSY particles and the
transverse momentum imbalance of the events. The four-momenta
of the two jets as well as the missing transverse momentum
${\vec{\mathrm E}}_{\mathrm T}^{\mathrm{miss}}$  may be used to compute
$\mathrm{M_R}$ and $\mathrm{R^2}$, defined as
\begin{eqnarray}
 \label{eq:MRstar}
 \mathrm{M_R} &\equiv&
 \sqrt{
(|\vec{p}_{{\mathrm j}_{1}}|+|\vec{p}_{{\mathrm j}_{2}}|)^2 -({p}^{{\mathrm j}_1}_z+{p}^{{\mathrm j}_2}_z)^2}\\
\mathrm{R^2} &\equiv&  \frac{{\mathrm E}_{\mathrm T}^{\mathrm{miss}}(p_{\mathrm{T}}^{{\mathrm j}_1}+p_{\mathrm{T}}^{{\mathrm j}_2}) -
 \vec{\mathrm E}_{\mathrm T}^{\mathrm{miss}} {\mathbf \cdot}
 (\vec{p}_{\mathrm{T}}^{\,{\mathrm j}_1}+\vec{p}_{\mathrm{T}}^{\,{\mathrm j}_2})}{4M_R^2}
\end{eqnarray}
where $\vec{p}_{{\mathrm j}_i}$, $\vec{p}_{\mathrm{T}}^{\,{\mathrm j}_i}$, and
$p^{{\mathrm j}_i}_z$ are the momentum of the ith-jet, its transverse
component, its longitudinal component, respectively, while ${\mathrm
  E}_{\mathrm T}^{\mathrm{miss}}$ and $p_{\mathrm{T}}^{{\mathrm j}_i}$ are the magnitude of
  $\vec{\mathrm E}_{\mathrm T}^{\mathrm{miss}}$ and $\vec{p}_{\mathrm{T}}^{\,{\mathrm
      j}_i}$, respectively. 

The search for SUSY is carried out on the events selected by a set of
criteria summarized in Table~\ref{tab:boxDef}. The events are detected by a set of dedicated
triggers, consisting of a loose selection on $\mathrm{M_R}$ and
$\mathrm{R^2}$. The events are also required to satisfy a requirement of two
jets in the central part of the detector. The trigger efficiency is measured
to be $(95 \pm 5)\%$.

Jets are reconstructed by clustering the Particle Flow (PF)~\cite{PF1,PF2} candidates with the {\sc
 FastJet}~\cite{fastjet} implementation of the anti-k$_{\mathrm
 T}$ algorithm~\cite{antikt} with the jet size set to $R=0.5$. We
select events containing at least two jets with $p_{\mathrm{T}}>80$ GeV and
$|\eta|<2.4$. For each event, the $\vec{\mathrm E}_{\mathrm T}^{\mathrm{miss}}$ and the
four-momenta of all the jets with $p_{\mathrm T} > 40$ GeV and $|\eta|<2.4$ are
used to compute the razor variables.

The medium working point of the combined secondary vertex
algorithm~\cite{btag8TeV} is used for jet b-tagging. 
Events without at least one b-tagged jet are discarded, a criterion
motivated by the expectation of a light top or bottom squark accessible
at LHC from naturalness considerations. A tighter
requirement ($\geq 2$ b-tagged jets) is imposed on events with less than four jets to reduce the
$\mathrm{Z}(\to\nu\bar\nu)$+jets background to a negligible level. 

\begin{table}[ht!]
\begin{center}
\begin{tabular}{|c|c|c|c|c|}
\hline
 \multicolumn{5}{|c|}{Requirements}\\ \hline
Box & lepton & b-tag & kinematic & jet \\\hline
\multirow{2}{*}{MultiJet} & \multirow{2}{*}{none} & \multirow{2}{*}{$\geq 1$
b-tag} & ($\mathrm{M_R} > 400$ GeV and $\mathrm{R^2} > 0.25$) and & \multirow{2}{*}{$\geq 4$ jets}\\
& & & ($\mathrm{M_R} >450$ GeV or $\mathrm{R^2} > 0.3$) & \\\hline
\multirow{2}{*}{2b-tagged jet} & \multirow{2}{*}{none} & \multirow{2}{*}{$\geq 2$
b-tag} & ($\mathrm{M_R} > 400$ GeV and $\mathrm{R^2} > 0.25$) and &
\multirow{2}{*}{$2$ or $3$ jets}\\
& & & ($\mathrm{M_R} > 450$ GeV or $\mathrm{R^2} > 0.3$) & \\\hline
\end{tabular}
 \caption{Kinematic and multiplicity requirements defining the two
   zero-lepton razor boxes. }
\label{tab:boxDef}
\end{center}
\end{table}

\section{Modeling the standard model backgrounds}
Under the hypothesis of no contribution from new physics processes,
the event distribution in the ($\mathrm{M_R}$, $\mathrm{R}^2$) plane
can be described by the sum of the weak vector boson plus jets
production ($\mathrm{V}$+jets where $\mathrm{V} =\mathrm{W},\mathrm{Z}$) and the top
quark-antiquark and the top single-quark production, generically referred to
as the $t\bar{t}$ contribution.

Based on the study of the data collected at $\sqrt{s}=7$ TeV and the
corresponding MC samples~\cite{razorPRL,razorPRD}, the two-dimensional
probability density function
$\mathrm{P_{SM}}(\mathrm{M_R},\mathrm{R^2})$ of each SM process is
found to be well described by the template function:
\begin{equation}
 f(\mathrm{M_R},\mathrm{R^2}) =  [b(\mathrm{M_R}-\mathrm{M_R^0})^{1/n}(\mathrm{R^2}-\mathrm{R^2_0})
  ^{1/n}-1]e^{-bn(\mathrm{M_R}-\mathrm{M_R^0})^{1/n}(\mathrm{R^2}-\mathrm{R^2_0})
    ^{1/n}} .
\label{eq:razFun}
\end{equation}
where $b$, $n$, $\mathrm{M_R^0}$, and $\mathrm{R^2_0}$ are free
parameters of the background model. The shape of the template function
is determined through an extended maximum likelihood (ML) fit to the data. The
template function is found to adequately describe the standard model
background in each of the boxes, for each b-tagged jet multiplicity.

\begin{figure}[htb]
\centering
\includegraphics[width=0.45\linewidth]{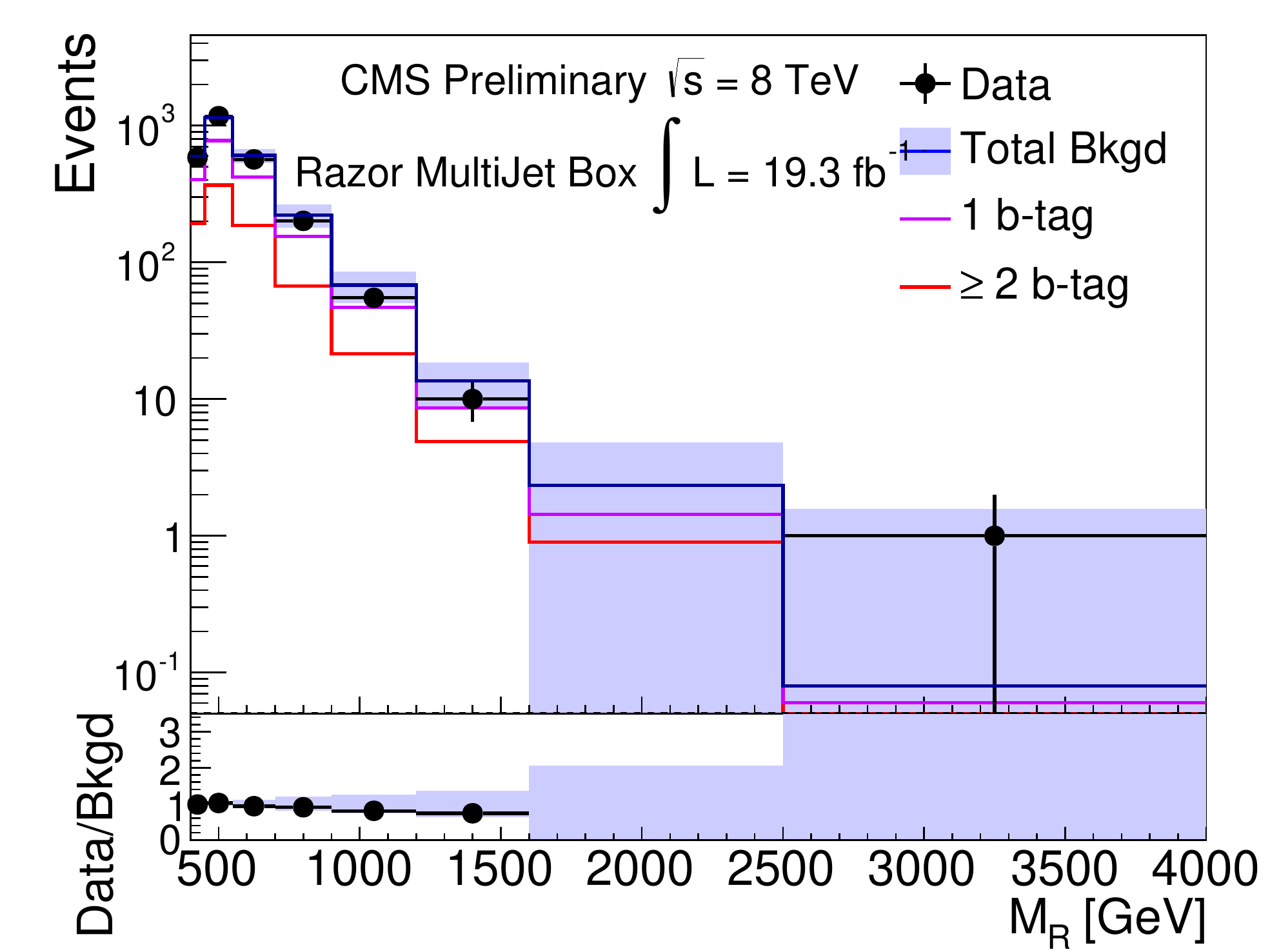}
\includegraphics[width=0.45\linewidth]{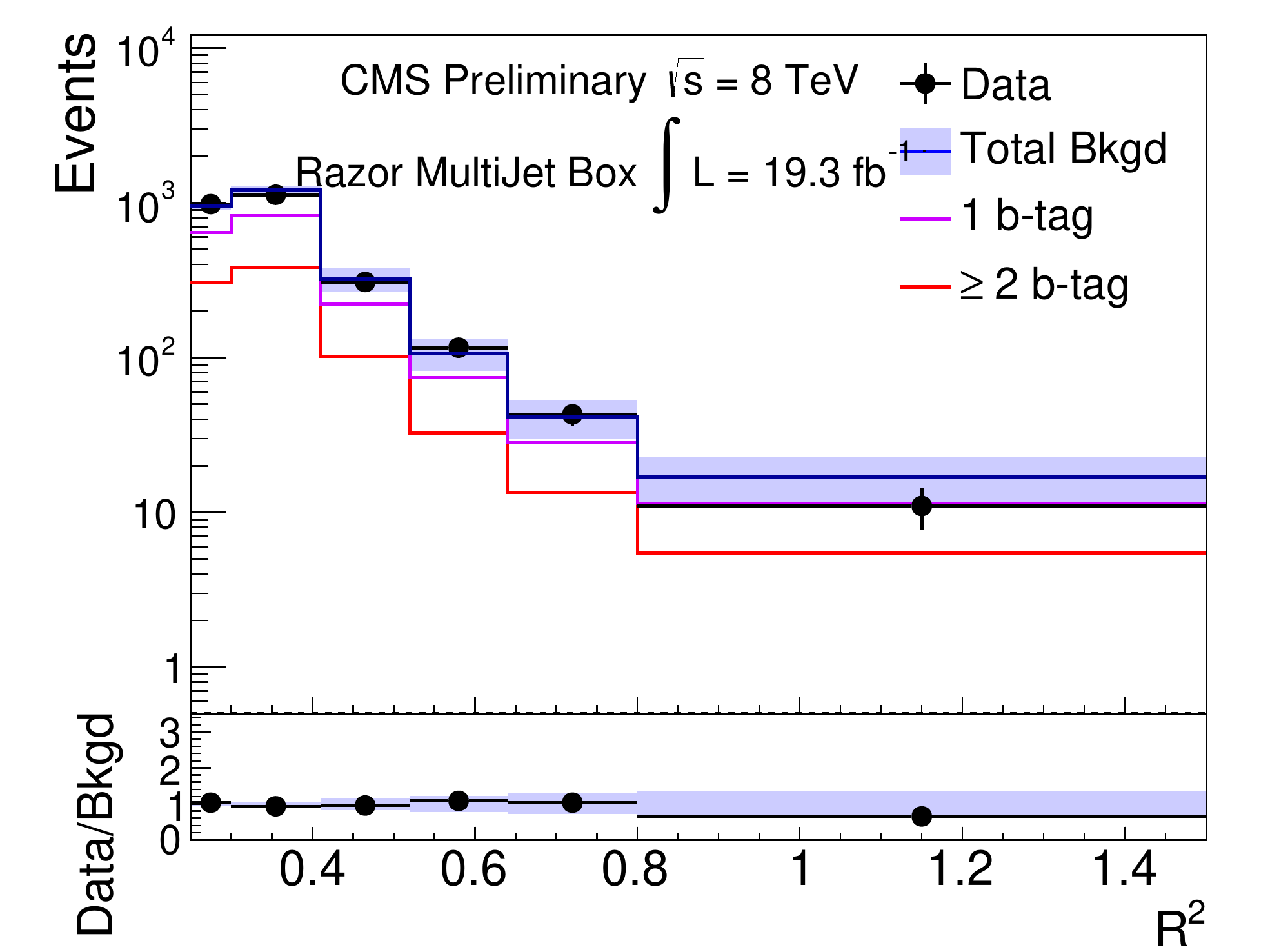}
\includegraphics[width=0.45\linewidth]{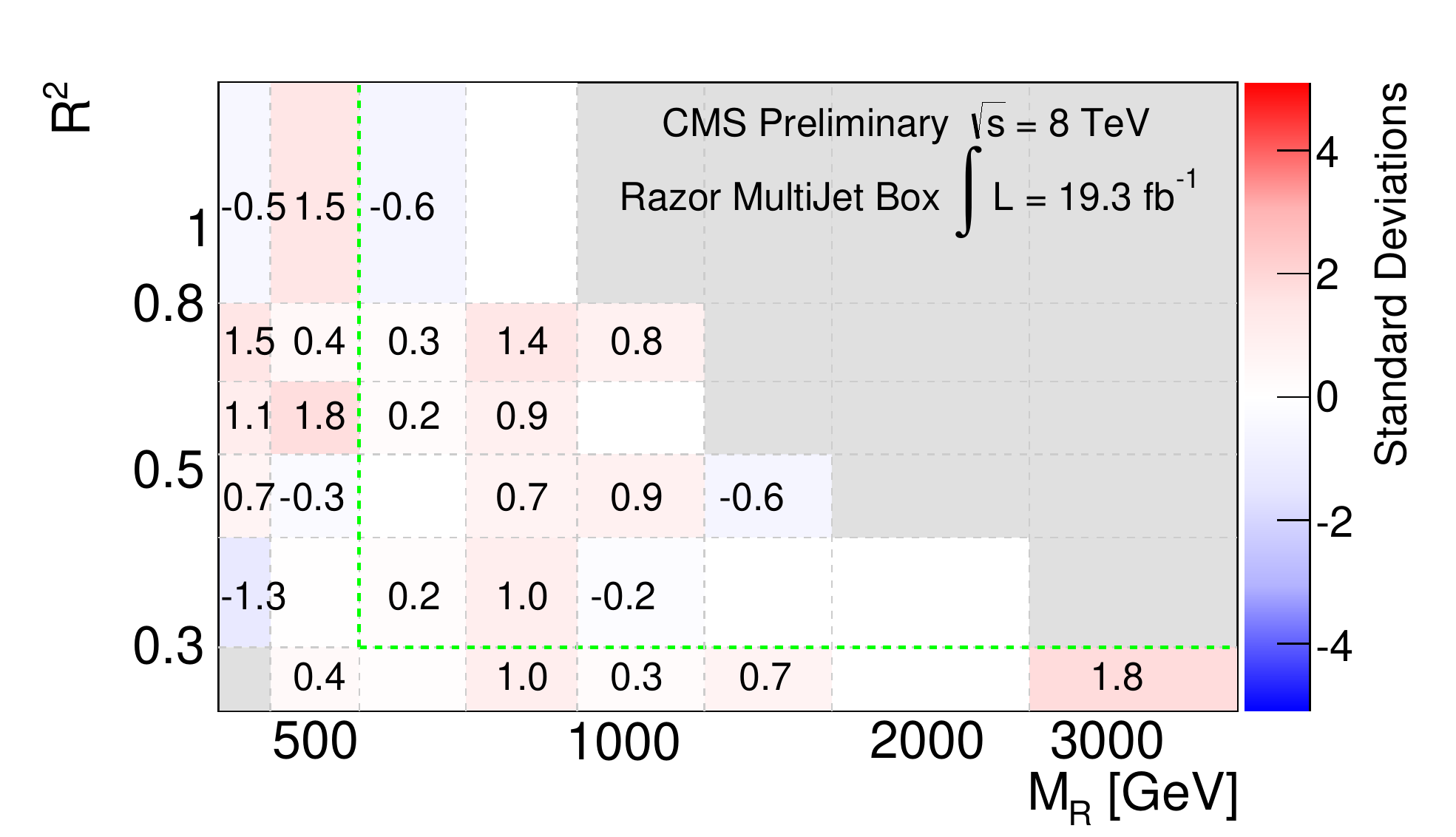}
    \caption{(Left and middle) Projection of the sideband fit result in the
      MultiJet box on $\mathrm{M_R}$ and $\mathrm{R^2}$, respectively. The solid line and the filled band represent the total
  background prediction and its uncertainty. The points and the band
  in the bottom panel represent the data-to-prediction ratio and the
  prediction uncertainty, respectively.  (Right) Comparison of the expected background and the observed yield
  in the MultiJet box in terms of a two-sided p-value. The
  p-value is translated into the corresponding number of standard
  deviations, quoted in each bin and represented by the bin-filling
  colors.}
    \label{fig:mrrsqData}
\end{figure}

The background shape parameters are estimated from the events in two
sidebands at low $\mathrm{M_{R}}$ ($\mathrm{M_R}<550$) and at low
$\mathrm{R^{2}}$ ($\mathrm{R^2}<0.3$). This shape is then used to derive the
background prediction in the signal-sensitive region along with a systematic
uncertainty associated to the background shape. Figure~\ref{fig:mrrsqData}
illustrates the agreement between the observation and the background
prediction in the MultiJet box. No significant deviation is observed.

\section{Interpretation and conclusions}

\begin{figure}[htb]
\centering
\includegraphics[width=0.3\linewidth]{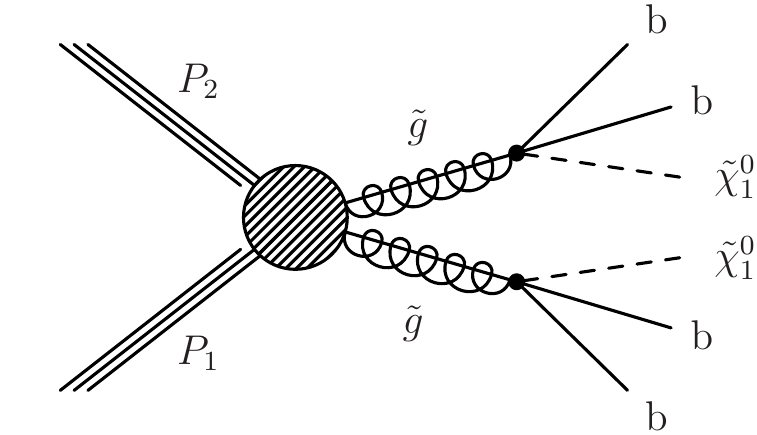}~~~~~~~~~~
\includegraphics[width=0.4\linewidth]{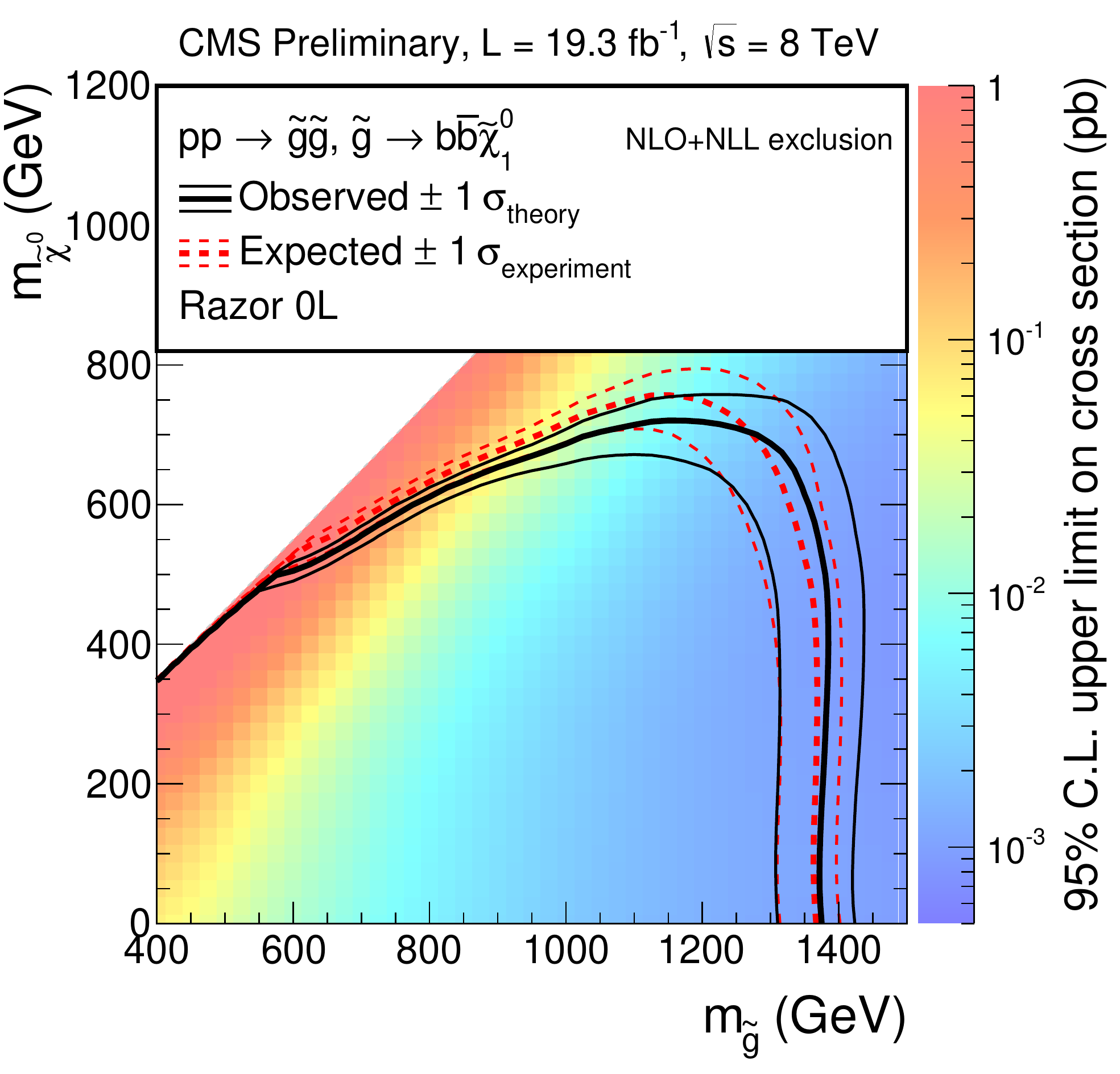}
    \caption{(Left) A diagram displaying the event topology of a
      gluino-mediated SUSY simplified model in the multi-bottom final state. (Right) Interpretation of the inclusive search with razor variables
  in the context of a gluino-mediated model. The color coding denotes the
  observed 95\%~C.L. upper limit on the SUSY signal cross section. The dashed
  and solid lines represent the expected and observed exclusion
  contours at 95\%~C.L., respectively. }%The dashed contours around the
  %expected limit and the solid contours around the observed one
  %represent the impact of the theoretical uncertainty in the cross
  %section and the combination of the statistical and experimental
  %systematic uncertainties, respectively.}
    \label{fig:signal}
\end{figure}

An interpretation of the results in a representative SUSY simplified
model is shown in Figure~\ref{fig:signal}. The model topology consists of
gluino pair-production, in which each gluino decays to a bottom quark,
a bottom antiquark and a LSP. Events for this SUSY simplified model
are generated with the {\sc MadGraph v5} simulation~\cite{Alwall:2014hca}, while the SUSY
particles are decayed and the event is showered in the {\sc Pythia v6}
simulation code~\cite{pythia}, before being processed through a
fast simulation of the CMS detector~\cite{fastsim}. The SUSY
particle production cross sections are calculated to next-to-leading
order (NLO) and next-to-leading-logarithm (NLL)
accuracy~\cite{NLONLL1,NLONLL2,NLONLL3,NLONLL4,NLONLL5}, assuming the
decoupling of the other SUSY partners. The NLO+NLL cross section and
the associated theoretical uncertainty~\cite{NLONLLerr} are taken as a
reference to derive exclusion limits on SUSY particle masses.

We carried out a search for supersymmetric particles using proton-proton
collision data collected by CMS at $\sqrt{s} = 8$ TeV. The dataset
size corresponds to an integrated luminosity of 19.3 fb$^{-1}$. We
analyzed events with at least two jets, at least one of which is
identified as a b-tagged jet, and study the event distribution in the
($\mathrm{M_R}$, $\mathrm{R}^2$) plane. No significant excess was observed over the standard model background
expectations, derived from a fit to the data distribution in
low-$\mathrm{M_R}$ and low-$\mathrm{R}^2$ sidebands.
The search results were translated into at 95\% confidence level
exclusion limits on the masses of the gluino and the LSP, in the
context of a simplified natural SUSY model.  For a LSP mass of
$100$~GeV, the pair production of gluinos in a multi-bottom final state was excluded for gluino masses up to $1375$~GeV.

\end{document}